\documentclass[aps,
superscriptaddress,
showpacs,
preprintnumbers,
nofootinbib,
nobibnotes,
bibnotes,
amsmath,
amssymb,
twocolumn,
prd,
floatfix,
]{revtex4-2}
\usepackage[utf8]{inputenc}
\usepackage{array}
\usepackage{subcaption}
\usepackage[normalem]{ulem}
\usepackage{xcolor}
 \usepackage{booktabs}

\newcolumntype{L}[1]{>{\raggedright\let\newline\\\arraybackslash\hspace{0pt}}m{#1}}
\newcolumntype{C}[1]{>{\centering\let\newline\\\arraybackslash\hspace{0pt}}m{#1}}
\newcolumntype{R}[1]{>{\raggedleft\let\newline\\\arraybackslash\hspace{0pt}}m{#1}}

\newcommand{\comment}[1]{}

\newcommand{\tr}{{\rm Tr}}
\newcommand{\ba}{\begin{eqnarray}}
\newcommand{\ea}{\end{eqnarray}}

\usepackage{graphicx}
\usepackage{dcolumn}
\usepackage{bm}
\usepackage{amsfonts}
\usepackage{natbib}
\usepackage{xcolor}
\usepackage{hyperref}
\usepackage{dcolumn}
\usepackage{subcaption}

\graphicspath{{figs/}}

\newcommand\regensburg{Fakult\"at f\"ur Physik, Universit\"at Regensburg, Universit\"atsstra{\ss}e 31, 93040 Regensburg, Germany}

\newcommand\mib{Dipartimento di Fisica, Universit\`a di Milano-Bicocca, Piazza della Scienza 3, I-20126 Milano, Italy}
\newcommand\infn{INFN, Sezione di Milano-Bicocca, Piazza della Scienza 3, I-20126 Milano, Italy}

\def\HVPlat{RBC:2018dos,Giusti:2019xct,Borsanyi:2020mff,Lehner:2020crt,Wang:2022lkq,Aubin:2022hgm,Ce:2022kxy,ExtendedTwistedMass:2022jpw,RBC:2023pvn,Kuberski:2024bcj,Boccaletti:2024guq,Spiegel:2024dec,RBC:2024fic,Djukanovic:2024cmq,ExtendedTwistedMass:2024nyi,MILC:2024ryz,Bazavov:2024eou}

\def\ILT{Hansen_2017,Bulava:2019kbi,Hansen:2019idp,Bailas:2020qmv,Gambino:2020crt,Bulava:2021fre,DelDebbio:2022qgu,Boito:2022njs,Gambino:2022dvu,ExtendedTwistedMassCollaborationETMC:2022sta,Barone:2023tbl,Bergamaschi:2023xzx,Bennett:2024cqv,Bruno:2024fqc,Patella:2024cto,Davier:2023cyp,DelDebbio:2024lwm,Tsuji:2026zku,Giusti:2026mcy,Abbott:2026wdw,Jay:2026qoh,Lupo:2026vdj,Abbott:2026juo,Lehner:2026jdg}

\def\TauData{OPAL1999Spectral,CLEO2000PiPi,ALEPH2005Final,Belle2008PiPi,Belle2010ThreeHadron,ALEPH2014Update,BaBar2018KK,Lusiani2019BaBar}

\def\DispHVP{Davier:2023fpl,Keshavarzi:2019abf,DiLuzio:2024sps,Leplumey:2025kvv,Bryzgalov:2026yaz}

\begin{document}
\title{\boldmath Reconstructing the hadronic vacuum polarization above a timelike cut in lattice QCD}
\author{M.~Bruno}\affiliation{\mib}\affiliation{\infn}
\author{C.~Lehner}\affiliation{\regensburg}
\author{J.~Parrino}\email{julian.parrino@ur.de}\thanks{Corresponding author}\affiliation{\regensburg}
\author{H.-Y.~Wang}\affiliation{\regensburg}
\noaffiliation

\collaboration{RBC and UKQCD Collaborations}

\date{\today}



\begin{abstract}
Hadronic $\tau$ decays are a possible alternative to $e^+e^-$ scattering data for a high-precision determination of the hadronic vacuum polarization (HVP) contribution from experimental data. For the former one must supplement the part of the spectrum above the $\tau$ mass, naturally not present in the experimental data. In this work, we calculate it from first-principles lattice QCD simulations, in the absence of isospin-breaking effects. More specifically, we provide a strategy to reconstruct the HVP contribution above a certain timelike energy cut from the Euclidean vector-vector correlator. This poses a natural inverse problem, related to the more general inverse Laplace transform for Euclidean correlators. In our work, for the systematic uncertainty due to the approximate reconstruction of the smeared spectral density we use the experimental data from $e^+e^-$ scattering. In addition, to further stabilize the inverse problem we subtract all states below 1 GeV. We present results for the intermediate-distance window observable for several energy cuts, including the $\tau$ mass, calculated on four physical point ensembles generated by the RBC/UKQCD collaborations.
\end{abstract}

\maketitle

\section{Introduction}
In the current standard model prediction of the anomalous magnetic moment of the muon \cite{Aliberti:2025beg}, the hadronic vacuum polarization (HVP) contribution amounts to the largest part of the uncertainty. 
In the first white paper from the muon $g-2$ theory initiative~\cite{Aoyama:2020ynm} the estimate of the HVP was dominated by the data-driven approach based on $e^+e^-$ scattering experiments \cite{\DispHVP}. Thanks to significant progress and increased accuracy in first-principles lattice QCD calculations and due to more tensions developing on the data-driven side, in particular in the two-pion channel, the current estimate of the recent update~\cite{Aliberti:2025beg} is based mostly on lattice QCD results ~\cite{\HVPlat}.
From a field theoretical approach the HVP can be equivalently defined as an integral over timelike or Euclidean domains \cite{Bernecker:2011gh} making it directly accessible to lattice QCD simulations, which are naturally defined with Euclidean metric. 
An important step that led not only to higher precision but also to a better scrutiny of the HVP contribution to $a_\mu$ was the splitting of the aforementioned integral in different windows \cite{RBC:2018dos}: in fact short, intermediate and long distances are affected differently by the typical lattice systematic errors, such as discretization or finite-volume effects. By targeting individual windows through separate and dedicated calculations, it was possible to reduce the overall uncertainty and reach a strong agreement among different lattice collaborations.

Beyond the comparison among different Lattice groups, window quantities can also be straightforwardly calculated using timelike data thereby offering additional intermediate observables to cross-check with the dispersive method. In this way, a  discrepancy between the two approaches was established.

In parallel to the computation from electron-positron data there is an ongoing effort for the determination of the HVP contribution which starts from hadronic $\tau$ decay data \cite{\TauData}.
To connect the weak decay matrix element to the electromagnetic vector correlator used in the determination of the HVP it is necessary to estimate the isospin breaking corrections in this process. A strategy to compute these effects in lattice QCD has been recently laid out  in Ref.~\cite{Bruno:2026yba}. A complementary analysis in the dispersive approach is carried out in Ref.~\cite{Allen:2026iad}.
The method developed in Ref.~\cite{Bruno:2026yba} is fully inclusive\footnote{In contrast to that the exclusive method, used in the WP2025 \cite{Aliberti:2024fpq}, only replaces the $\pi^+\pi^-$ channel of the $e^+e^-$ scattering dataset by the $\pi^+\pi^0$ channel of the $\tau$ decay dataset after an isospin rotation.} and utilizes the complete $\tau$ decay dataset, which is only available up to the mass of the $\tau$ lepton $m_\tau$ due to kinematic constraints. The missing contribution above $m_\tau$ therefore needs to be calculated by different means.

In this work, taking into account the need emerging from the method based on $\tau$ data together with the necessity to better pin down the discrepancy between the lattice QCD and $e^+e^-$ data-driven determination, we define a new observable, $a_\mu^{>}(\omega_{\text{cut}})$, as the contribution to the HVP above the (timelike) energy threshold $\omega_{\text{cut}}$ of the underlying spectral density. For the $\tau$ decay project the most natural starting point is $a_\mu^{>}(\omega_{\text{cut}}=m_\tau)$, but eventually, one may consider lower values of $\omega_{\text{cut}}$ to possibly reduce some systematic uncertainties from the experimental dataset. A calculation of $a_\mu^{>}(\omega_{\text{cut}}=m_\tau)$ in perturbative QCD (pQCD) has already been carried out in Ref. \cite{Allen:2026iad}. 

Due to the presence of a step function, a computation in lattice QCD constitutes an ill-posed problem since it requires an inverse Laplace transform. However, there have been multiple developments within the lattice community to address this issue~\cite{\ILT}.
One possible approach is to calculate the smeared density as a linear combination of different timeslices of the relevant correlation function. Among the several variants studied in the literature, a widely used approach is the Hansen-Lupo-Tantalo (HLT) method~\cite{Hansen:2019idp} applied to determine the smeared R-ratio \cite{DeSantis:2022qht,ExtendedTwistedMassCollaborationETMC:2022sta,Blum:2024hyr,ExtendedTwistedMass:2025tpc,Margari:2026him} and the inclusive hadronic $\tau$ decay rate, relevant in the determination of CKM matrix elements~\cite{Evangelista:2023fmt,DiCarlo:2026kpv}.
The research on smeared spectral densities is an active field of research with many recent developments \cite{Bresciani:2026kjv,Jay:2026qoh}.
In our work, we present the first calculation of $a_\mu^{>}(\omega_{\text{cut}})$ from lattice QCD. We start by considering a target kernel involving a smeared Heaviside function, rather than a sharp one, and derive the coefficients required in the reconstruction using the HLT method~\cite{Hansen:2019idp} with a different regulator~\cite{Bruno:2024fqc}.
Beyond the specific target quantity studied here, our work differs w.r.t.~the typical literature on the subject in two main aspects.  These aspects are very general and applicable to a vast variety of other studies.

Firstly, if the simulations are tuned to the physical point and isospin breaking effects can be ignored, available experimental data can be used in the estimate of the systematic uncertainty. Even in the case of the R-ratio where significant tensions are present, for the sole purpose of addressing the difference between the approximated and the true kernel, the current precision is sufficient. This may be true in several other contexts as well.

Secondly, another important aspect in our approach is that by building on Ref.~\cite{Bruno:2020kyl} we actively study the impact of the subtraction of the lowest lying states from the lattice correlator on the stability of the reconstruction. Evidently this is driven by the interplay between the density and the kernel, and while in general a direct knowledge of the spectrum leads to an obvious solution of the inverse problem up to that scale~\cite{Bruno:2020kyl}, it can significantly improve the stability of the reconstruction, even for highly energetic kernels as the ones studied here.
Both aspects lead to a better control of the overall error and in general to a more precise outcome.

We perform an analysis on four physical point ensembles generated by the RBC/UKQCD collaborations and extensively used in the determination of the HVP window observables. In particular, by re-using the knowledge of the spectrum in the I=1 vector channel from the long-distance window study of Ref.~\cite{RBC:2024fic} we subtract all states below 1 GeV. We extrapolate to the continuum limit using three lattice spacings.

This manuscript is organized as follows: In section \ref{sec:method}, we define the target observable, $a_\mu^{>}(\omega_{\text{cut}})$ through a cut in the underlying spectral density and discuss its computation in pQCD. Section \ref{sec:inverse_problem} describes the inverse problem, which has to to be dealt with, when computing $a_\mu^{>}(\omega_{\text{cut}})$ in lattice QCD and describes how we estimate the systematic uncertainty. In section \ref{sec:recon_lat}, we apply the method defined in the previous sections to lattice QCD data. Further analysis of the results obtained from the reconstruction on different lattice QCD ensembles is described in section \ref{sec:analysis_lat}, also providing the extrapolation to the continuum. We conclude in section \ref{sec:discussion} with a discussion of the results.
\section{Hadronic vacuum polarization above a timelike threshold}
\label{sec:method}
The hadronic vacuum polarization (HVP) contribution to the anomalous magnetic moment of the muon in the time-momentum representation is given by \cite{Bernecker:2011gh}
\ba 
\label{eq:a_mu_TMR}
    a^{\text{HVP}}_\mu = \int_0^\infty dt\ G(t) f(t,m_\mu)
\ea 
where 
\ba
\label{eq:spatsummedcor}
G(t)\ = \frac{1}{3} \sum_{k=1,2,3} \int d^3\vec{x} \langle j_k(t,\vec{x})j_k(0) \rangle_{\text{QED}+\text{QCD}},
\ea
is the spatially summed correlator of two electromagnetic vector currents $j_\mu(x)=i\sum_{f}Q_f\bar{\psi}_f(x)\gamma_\mu \psi_f(x)$, for quark flavors $f=\{u,d,s,c,b,t\}$ with charges $Q_f=\{\frac{2}{3},-\frac{1}{3},-\frac{1}{3},\frac23,-\frac13,\frac23 \}$ and $f(t,m_\mu)$ is a known kernel function. From Eq.~\eqref{eq:a_mu_TMR} the HVP contribution can be calculated in Euclidean space on the lattice, where the vector correlator \eqref{eq:spatsummedcor} needs to be calculated in QED+QCD.

In Ref.~\cite{Bruno:2026yba}, a strategy for a fully inclusive determination of the HVP from $\tau$-decay data is proposed. In this strategy, lattice QCD input is used to perform the isospin rotation necessary to connect the charged hadronic decay of the $\tau$ to the charge neutral electromagnetic hadronic decay. In this regard, it is instructive to define the charged \footnote{Eq.~\eqref{eq:def_GW_correlator} is defined with a minus sign matching the convention of Ref.~\cite{Bruno:2026yba}, in contrast to Eq.~\eqref{eq:spatsummedcor}, which is matched to Ref.~\cite{RBC:2024fic}}
\ba 
\label{eq:def_GW_correlator}
\nonumber
    G^W(t) &= &\,  -\frac13 \sum_{k=1,2,3} \int d^3 \vec x \langle j^{+}_k(t,\vec x) j^{-}_k(0) \rangle_{\text{QED}+\text{QCD}},
    \\
    \label{eq:GWtmu}
\ea
for the charged vector currents
\begin{equation}
    j^+_\mu(x) =  \, \frac{1}{\sqrt 2} (\bar d \gamma_\mu u)(x) \,, \quad j^-_\mu(x) = \frac{1}{\sqrt 2} (\bar u \gamma_\mu d)(x) \,. \\
\end{equation}
In the isospin symmetric limit, the correlator $G^W(t)$ is therefore equal to the isovector component of the correlator of the electromagnetic vector current \eqref{eq:spatsummedcor}. Any difference between these stems from isospin breaking corrections, which we write as $O(\alpha)$ in the following. In the spectral representation, we express Eq.~\eqref{eq:def_GW_correlator} as
\ba 
\label{eq:def_GW}
G^W(t) = \int_0^\infty  d\omega \, \omega^2 \, e^{-\omega t} \, \rho^{\rm{I=1}}(\omega^2)+O(\alpha),
\ea 
with only isovector states contributing to the spectral density $\rho^{\rm{I=1}}(\omega^2)$ and isospin breaking corrections denoted by $O(\alpha)$.
The spectral density $\rho^{\rm{I=1}}(\omega^2)$ is related to the R-ratio by
\ba 
\rho^{\rm{I=1}}(\omega^2)= \frac{R^{\rm{I=1}}(\omega^2)}{12\pi^2}, 
\ea
\ba 
R^{\rm{I=1}}(\omega^2):= \frac{\sigma(e^+e^- \rightarrow \text{hadrons (I=1)})}{4\pi\alpha(\omega^2)^2/(3\omega^2)}.
\ea 
Ref.~\cite{Bruno:2026yba} provides a strategy to address the $O(\alpha)$ effects in Eq.~\eqref{eq:def_GW}. These effects constitute a correction to the experimental $\tau$ data, which is then used to determine the HVP. 

In this work, we want to focus on another aspect that needs to be accounted for, if one wants to use the full $\tau$ decay data. The experimental data for the R-ratio from hadronic $\tau$ decay \cite{\TauData} is only available up to a cut in the energy spectrum given by the mass of the $\tau$-lepton, $m_\tau$. In order to provide a determination of the HVP contribution from this dataset, one needs to estimate the contribution above $m_\tau$ by other means. This issue has also recently been discussed in Ref.~\cite{Allen:2026iad}. Here, we want to point out that by considering cuts in the energy spectrum below $m_\tau$, the systematic uncertainty from the $\tau$ decay dataset may be further reduced.
To this end, we define the isovector correlator above a timelike energy cut $\omega_{\text{cut}}$ as 
    \begin{equation}
        \label{eq:def_GWlarger}
        G^{>}(t,\omega_{\text{cut}}) = \int_0^\infty d\omega \, \omega^2 \, e^{-\omega t} \, \rho^{\rm{I=1}}(\omega^2) \, \theta(\omega - \omega_{\text{cut}}). 
    \end{equation}
By inserting Eq.~\eqref{eq:def_GWlarger} into Eq.~\eqref{eq:a_mu_TMR}, we obtain the HVP contribution above a timelike cut $a_\mu^{>}(\omega_{\text{cut}})$. This idea is similar to the introduction of Euclidean time window contributions \cite{RBC:2018dos}, which separate different systematic uncertainties of lattice QCD calculations. Combining both windowing procedures, we define the HVP window contribution above a timelike cut as
\begin{equation}
\label{eq:a_mu_larger_win}
    a_\mu^{>,\text{win}}(\omega_{\text{cut}}) = \int_0^\infty dt \, f(t,m_\mu) \, G^{>}(t,\omega_{\text{cut}}) \, \Theta^\text{win}(t) \,.
\end{equation}
Here, $\Theta^{\text{win}}(t)$ is a smooth window function
\begin{equation}
    \Theta^{\text{win}}(t) = \frac12\left(\tanh\left(\frac{t-t_0}{\Delta_{\text{win}}}\right) - \tanh\left(\frac{t-t_1}{\Delta_{\text{win}}}\right)\right).
\end{equation}
We consider the canonical choice of $t_0,t_1,\Delta_{\text{win}}$ for the  short-distance (SD), intermediate-distance (ID), and long-distance (LD) windows \cite{Aliberti:2025beg}.

In the timelike picture, the window contribution to $a_\mu$ is given by 
\ba 
\nonumber
a_\mu^{>,\text{win}}(\omega_{\text{cut}})&=&8\alpha^2\int_0^\infty   d\omega \Big[ \frac{K(\omega^2)}{\omega}\rho^{\rm{I=1}}(\omega^2)\\
 && \hspace{1.3cm} \tilde{\Theta}^{\text{win}}(\omega^2)  \theta(\omega - \omega_{\text{cut}}) \Big]
\label{eq:window_dispersive_ahvp}
\ea 
where the kernel $K(\omega^2)$ and the function $\tilde{\Theta}^{\text{win}}(\omega^2)$ can be found in Ref. \cite{Colangelo:2022vok}. Eq.~\eqref{eq:window_dispersive_ahvp} states both windows explicit: In Euclidean time via $\tilde{\Theta}^{\text{win}}(\omega^2)$ and in timelike energy by $\theta(\omega - \omega_{\text{cut}})$ \footnote{If no cut is applied both window functions are exactly one}.

\begin{table*}
\caption{Phenomenological contributions and pQCD estimates in units of $10^{-10}$. For the phenomenological contributions the errors are obtained from statistical as well as channel-specific uncertainties ($1.4\%$ for the $\pi^+\pi^-$ channel and $3\%$ for the four-pion channels) assuming no correlation between channels.
For our pQCD results, uncertainties are computed from varying the renormalization scale $\mu \in [\sqrt{s}/2,2\sqrt{s}]$ and the strong coupling at the Z-pole mass $\alpha_s = 0.118\pm 0.003$, similar to Ref.~\cite{Harlander:2002ur}. The resulting error band in this procedure is not symmetric around the central value. We therefore quote upper and lower bounds of the theoretical error band. For comparison, we show the pQCD results from Ref.~\cite{Allen:2026iad}, which includes a model for duality violations and an estimate of the 5th order coefficient. The quoted results from Ref.~\cite{Allen:2026iad} marked with a $^*$ use a cut that is slightly below the $\tau$ mass at $\omega_{\text{cut}}=1.75699$ GeV and are therefore separated with a double horizontal line. The shift for the pQCD results between this lower cut and $m_\tau$ computed with 4th order pQCD is roughly about to the size of the uncertainty of the 5th order results.}
\label{tab:pheno_and_pertQCD}
  \begin{tabular}{|l|l|l|l|l|l|l|l|}
  \hline
   & $a_\mu^{>}(\omega_{\text{cut}}=m_\tau)$ &  $a_\mu^{>,SD}(\omega_{\text{cut}}=m_\tau)$ &  $a_\mu^{>,ID}(\omega_{\text{cut}}=m_\tau)$ & $a_\mu^{>,LD}(\omega_{\text{cut}}=m_\tau)$  \\ \hline
  $\pi^+\pi^-$ & $0.237[0.009]$&$ 0.089[0.003] $& $0.138[0.005] $ &$  0.011   [0.001] $  \\ \hline
  $\pi^+\pi^-+2\pi^+ 2\pi^-+\pi^+\pi^- 2\pi^0$ & $7.871   [0.169] $& $3.462    [0.075]$ & $4.155     [0.089] $& $ 0.253   [0.005] $  \\ \hline
  pQCD (4th order)
  &$ 33.67   [^{+0.25}_{-0.58}]  $& $ 20.03     [^{+0.13}_{-0.25}]$ & $13.05     [^{+0.12}_{-0.31}]$ &   $0.59   [^{+0.01}_{-0.02}]$  \\ \hline \hline
  pQCD (5th order + DV) from \cite{Allen:2026iad} &$ ^*33.77     [0.57] $& $ ^*20.03     [0.19]$ & $ ^*13.12     [0.35]$ &   $ ^*0.602   [0.035]$  \\ \hline
  \end{tabular}
\end{table*}
Below $\omega_{\text{cut}}=m_\tau$ the dominant hadronic contributions to the $I=1$ part of the R-ratio are the $\pi^+\pi^-$, $2\pi^+ 2\pi^-$ and $\pi^+\pi^- 2\pi^0$ channels \cite{Benton:2024kwp}. In the region above $m_\tau$, the contribution from these channels to $a_\mu^>(\omega_{\text{cut}})$ is largely suppressed. In this regime, a large number of additional hadronic decay channels need to be taken into account. This makes an exclusive dispersive treatment in this region very challenging \cite{Allen:2026iad}. In the dispersive evaluation of $a_\mu^{\text{HVP}}$, this is usually taken care of by resorting to a fully inclusive determination \cite{Aliberti:2025beg}. This however does not allow for a separation into isovector and isoscalar channels.

 We give the full contribution as well as the window observables for $a_\mu^{>}(\omega_{\text{cut}}=m_\tau)$ in Table \ref{tab:pheno_and_pertQCD} for the $\pi^+\pi^-$ channel and the $\pi^+\pi^-$, $2\pi^+ 2\pi^-$ and $\pi^+\pi^- 2\pi^0$ channels combined using data from the BaBar experiment \cite{BaBar:2012bdw,BaBar:2012sxt,BaBar:2017zmc}. Note that the $\pi^+\pi^-$ contribution, which currently exhibits a $3-4\sigma$ tension between different datasets \cite{Aliberti:2025beg}, is almost negligible for the determination of  $a_\mu^{>}(\omega_{\text{cut}}=m_\tau)$.

In contrast to the data-driven evaluation of $a_\mu^{>}(\omega_{\text{cut}})$, it is straight-forward to determine this contribution in perturbative QCD. Here, it is however important to note that the region $\sqrt{s}>m_\tau$ is just barely at the threshold, where perturbative QCD can be safely applied to compute the R-ratio. For energies $\omega>3$ GeV violations of the quark hadron duality (DV) are mostly negligible. However at $\omega \sim m_\tau$, the effect of DVs in the  R-ratio is very significant \cite{Boito:2025qfz}.

In perturbative QCD (pQCD) up to fourth order, the R-ratio for massless 3 flavour QCD below the charm threshold $3.7$ GeV is written as 
\ba 
\nonumber
R(\omega^2)&=&3\sum_{f=u,d,s} Q_f^2\Big[1+\Big(\frac{\alpha_s(\omega^2)}{\pi}\Big)+r_2\Big(\frac{\alpha_s(\omega^2)}{\pi}\Big)^2\\
&&\hspace{-0.3cm} +r_3 \Big(\frac{\alpha_s(\omega^2)}{\pi}\Big)^3+r_4 \Big(\frac{\alpha_s(\omega^2)}{\pi}\Big)^4 +O(\alpha_s^5) \Big]
\ea 
with the numerical coefficients $r_i$  computed a long time ago in Eq.~(5)  of Ref.~\cite{Baikov:2008jh}\footnote{An estimate of the fifth order coefficient is provided in Ref.~\cite{Herzog:2017ohr} with $50\%$ relative uncertainty.}. Here, the energy dependence of the running coupling $\alpha_s(\omega^2)$ is very mild. For massless quarks, the isovector part of the R-ratio is given by simply rescaling $R^{I=1}(\omega^2)=\frac{3}{4}R(\omega^2)$.
Our numerical results for $R^{I=1}(\omega^2)$ are obtained using the rhad software package \cite{Harlander:2002ur} and are given in table \ref{tab:pheno_and_pertQCD}.
Here, it is worth mentioning that since the kernel for the short-distance window contribution falls off very slowly, there is still a non negligible contribution from the region above 10 GeV.
In our calculation, we neglect the fifth order term and the effect of DVs on the results for the window contributions. Our results are however in agreement with the results of Ref.~\cite{Allen:2026iad}, which includes both effects.

\section{Defining the inverse problem}
\label{sec:inverse_problem}

\begin{figure}
    \centering
    \includegraphics[width=0.99\linewidth]{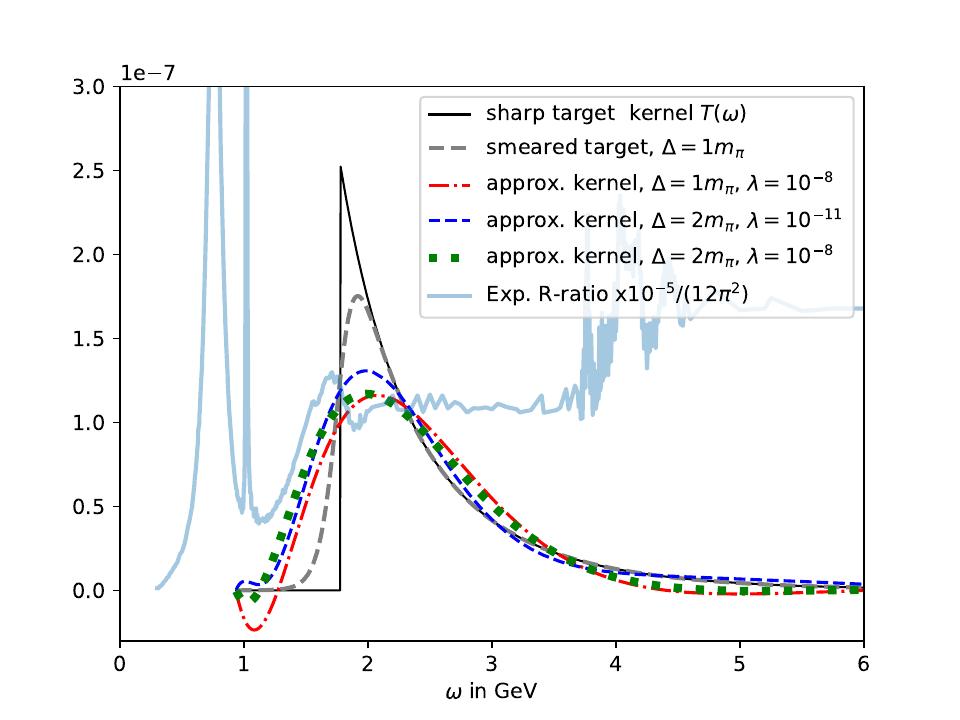}
    \caption{To illustrate the method from section~\ref{sec:method}, we compare the sharp target kernel for the total HVP contribution $T(\omega)$ (black line) and the smeared kernel (dotted grey line) to the approximated kernel $\widehat{T}_{\Delta,\lambda}(\omega)$ for multiple parameter pairs $[\lambda,\Delta]$ and using $p=2$. As a reference, we show the experimental R ratio from $e^+e^-$ scattering, multiplied with a scale factor for visibility. The coefficients of the approximated kernel are obtained using Eq.~\eqref{eq:q_coef_endpoint_constraint} with the endpoint constraint, $\widehat{T}_{\Delta,\lambda}(\omega_0)=0$.}
    \label{fig:kernel_mismatch}
\end{figure}
In this section, we want to describe the method for obtaining $a_\mu^{>,\text{win}}(\omega_{\text{cut}})$, defined in Eq.~\eqref{eq:window_dispersive_ahvp} from lattice QCD. We concentrate on the case of isospin symmetric QCD for now.
The isovector correlator $G^{I=1}(t)$ can be directly computed from lattice QCD. This correlator receives only a contribution from the connected Wick contraction 
\ba 
\nonumber
G^{I=1}(t) = \frac{1}{6} \sum_{k} \int d^3\vec x 
    \langle \tr \big[
     \gamma_{k} D^{-1} (x,t;0) \gamma_k D^{-1} (0;x,t)
    \big] \rangle , 
\ea
where the expectation value needs to be taken over QCD gauge configurations. Here, the sum is performed over $k=1,2,3$. $D^{-1}$ is the inverse of the Dirac operator for light quarks. 
Since the contribution $G^{>}(t)$, Eq.~\eqref{eq:def_GWlarger}, is defined with a cut in the timelike regime, obtaining $G^{>}(t)$ directly from lattice QCD poses an inverse problem.
In order to deal with this, we define the sharp target kernel by 
\ba
\label{eq:target_energy_kernel_integrated}
T(\omega)
=
8\alpha^2 K(\omega^2)\omega^{p-3}\tilde{\Theta}^{\text{win}}(\omega^2) \theta(\omega-\omega_{\text{cut}}).
\ea
Integrated with the spectral density and a compensating polynomial in $\omega$ this yields
\ba 
\label{eq:a_larger_with_target_kernel}
a_\mu^{>,\text{win}} = \int_0^\infty  d\omega \, T(\omega) \, \omega^{2-p} \rho^{I=1}(\omega^2),
\ea 
which is equivalent to Eq.~\eqref{eq:window_dispersive_ahvp}. Note how the additional factors $\omega^p$ cancel out in the final integrated observable. In practice we exploit this freedom and observe that absorbing a polynomial with $p=4$ in the basis functions leads to a more stable reconstruction.
In a practical lattice calculation only a smeared version of the sharp kernel can be estimated; to this end we introduce the target kernel
\ba
\label{eq:smeared_target_energy_kernel_integrated}
T_\Delta(\omega)
=
8\alpha^2 K(\omega^2)\omega^{p-3}\tilde{\Theta}^{\text{win}}(\omega^2) \theta_\Delta(\omega-\omega_{\text{cut}}),
\ea
defined from the smeared step function
\ba 
\theta_\Delta(\omega-\omega_{\text{cut}})= \frac{1}{2}\left[ 1+ \tanh\left(\frac{\omega-\omega_{\text{cut}}}{\Delta}\right)\right] \,,
\ea 
such that $\lim_{\Delta \to 0} T_\Delta(\omega) = T(\omega)$.
We want to estimate this kernel in the space spanned by the set of basis functions \footnote{The second exponential function in Eq.~\eqref{eq:basis_functions} accounts for the backwards propagating term in the correlator on a periodic lattice with temporal extent $T$. }
\ba
\label{eq:basis_functions}
\phi_t(\omega)=\omega^p\left[e^{-\omega t}+e^{-\omega(T-t)}\right].
\ea
Since we have a finite and discrete set of timeslices, the approximated kernel is then given by
\ba
\label{eq:approximated_kernel}
\widehat{T}_{\Delta,\lambda}(\omega)
=
\sum_n q_{n,\lambda}\phi_{t_n} (\omega) \,,
\ea
with real coefficients $q_{n,\lambda}$. In Fig.~\ref{fig:kernel_mismatch}, we display the sharp target kernel $T(\omega)$ together with the smeared target kernel. We compare these to the approximated kernel for multiple parameter pairs $[\lambda,\Delta]$.

By following Refs. ~\cite{BackusGilbert10.1111/j.1365-246X.1968.tb00216.x,Hansen:2019idp, Bruno:2024fqc}, we determine the coefficients $q_{n,\lambda}$ by minimizing the squared distance (in energy space) between the smeared target kernel \eqref{eq:smeared_target_energy_kernel_integrated} and the approximated kernel \eqref{eq:approximated_kernel} with an additional regulating term, proportional to $\lambda$, which tames the ill-posed nature of the problem. Additional details are given in Appendix \ref{sec:appendix_HLTimplementation}.\footnote{The lower bound for integration in energy space $\omega_0$, can be chosen at the lower end of the underlying spectrum, e.g. the two-pion threshold for the case of $\rho^{\rm I=1}(\omega^2)$.}
The reconstructed observable is calculated by combining the correlator with the numerical coefficients $q_{n,\lambda}$ according to
\ba
\nonumber
\Big(a_\mu^{>,\text{win}}(\omega_{\text{cut}})\Big)_{\Delta,\lambda} &=&\int_{\omega_0}^\infty  d\omega\, \widehat{T}_{\Delta,\lambda}(\omega) \, \omega^{2-p} \rho^{\rm I=1}(\omega^2) \\
\label{eq:HLT_recon_awin}
&=& \sum_n q_{n,\lambda} G^{{I=1}}(t_n) \,.
\ea
Its statistical uncertainty
\ba
\label{eq:GW_stat_unc}
 \sigma^{2}_{\text{stat}}
 =q^T~\mathrm{Cov}~ q \,,
 \ea
is easily obtained by propagating the errors from the covariance matrix of the lattice correlator $G^{I=1}(t)$, $\mathrm{Cov}$. The total uncertainty of the reconstructed quantity is given by $\sigma^2_{\text{tot}}= \sigma^{2}_{\text{stat}}+ \sigma^{2}_{\text{syst}}$ and typically the regulator $\lambda$, see Eq.~\eqref{eq:def_W_HLT}, is chosen in such a way that one balances the systematic uncertainties from the inexact inverse problem with the statistical noise, Eq.~\eqref{eq:GW_stat_unc}. 
The former may be defined from the integral
\ba
\label{eq:syst_uncertainty_estimate}
\sigma_{\text{syst}} = \left| \int_{\omega_0}^\infty d\omega
\left[
\widehat{T}_{\Delta,\lambda}(\omega)
-
T(\omega)
\right]\omega^{2-p} \rho^{I=1}(\omega^2) \right| \,.
\ea
Contrary to the typical approaches used in the literature where one looks for a stability in the $\lambda$ parameter at fixed $\Delta$, and later extrapolates $\Delta \to 0$ \cite{Evangelista:2023fmt,DiCarlo:2026kpv,ETMCtau24}
, in our work we leverage the experimental knowledge of the underlying spectral density to estimate directly Eq.~\eqref{eq:syst_uncertainty_estimate}. More specifically, we use the fact that $\rho(\omega^2)\geq \rho^{I=1}(\omega^2)\geq0$ and make use of the fully inclusive R-ratio data obtained from $e^+e^-$ scattering \cite{ParticleDataGroup:2026aaa} $R^{e^+e^-}(\omega^2)$. To account for possible deviations due to the differences in the experimental datasets, we add energy dependent Gaussian noise to $R^{e^+e^-}(\omega^2)$ with a maximum relative deviation of $5$\% and we vary the correlation length of this noise from $20$ to $200$ MeV. The systematic uncertainty that we adopt is defined by
\ba
\nonumber
\sigma_{\text{syst}}&=&\text{max}\Big\{\Big|
\int_{\omega_0}^\infty d\omega
\left[
\widehat{T}_{\Delta,\lambda}(\omega)
-
T(\omega)
\right] \omega^{2-p} \\
&&
\label{eq:syst_uncertainty_calc}
\hspace{1cm} \cdot \frac{R^{e^+e^-}(\omega^2)}{12\pi^2}  \Big[1+\sigma_{\text{noise}}(\omega)\Big]
\Big| \Big\},
\ea 
where only the maximum deviation over the set of different correlation lengths is considered. By considering Gaussian noise from the experimental R-ratio for the computation of the systematic uncertainty, we avoid to accidentally minimize it for the exact shape of the experimental R-ratio, an important consideration later on in the optimization of the choice of $\lambda$ and $\Delta$. 

Therefore in our approach, we expect to observe a plateau both w.r.t.~$\lambda$ and $\Delta$ thereby removing the need for an extrapolation, but leaving us with a leverage to find an optimal point in the $(\lambda, \Delta)$ plane where the total error is smallest.

We want to mention here that although more challenging, it is also possible to reconstruct $G^{>}(t')$ for each timeslice $t'$ using this method, by considering the target kernel
\ba
\nonumber
T_{\Delta}^{(G)}(\omega;t')=\omega^{p} \left[e^{-\omega t'}+e^{-\omega(T-t')}\right]\theta_\Delta(\omega-\omega_{\text{cut}}).
\\
\label{eq:target_kernel_integrand}
\ea
As proposed by Ref.~\cite{Bruno:2020kyl}, this may offer another approach to the determination of the spectral density itself.

\section{Reconstruction from lattice QCD data}
\label{sec:recon_lat}

\begin{table}
\caption{Parameters of the ensembles used in this work. All values are taken from Ref.~\cite{RBC:2024fic}.}
\label{tab:ensembles}
  \begin{tabular}{|l|l|l|l|l|l|l|l|l|}
  \hline
  ID & $a^{-1}$/GeV &  $L^3 \times T \times L_s/a^5$ & $m_\pi$/MeV & $m_K$/MeV & $m_\pi L$  \\ \hline
  48I & $1.7312(28)$ &  $48^3 \times 96 \times 24$ & $139.32(30)$ & $499.44(88)$ & 3.9  \\ \hline
  Ca & $1.7312(28)$ &  $64^3 \times 128 \times 24$ & $139.32(30)$ & $499.44(88)$ & 5.2  \\ \hline
  64I & $2.3549(49)$ &  $64^3 \times 128 \times 12$ & $138.98(43)$ & $507.5(1.5)$ & 3.8  \\ \hline
  96I & $2.6920(67)$ &  $96^3 \times 192 \times 12$ & $131.29(66)$ & $484.5(2.3)$ & 4.7  \\ \hline
  \end{tabular}
\end{table}

In this section, we want to give details of the computation carried out on four physical pion mass ensembles, generated by the RBC/UKQCD collaborations. Details of the ensemble parameters are given in Table \ref{tab:ensembles}. In the following, we will focus on the intermediate distance window quantity using the kernel given in Eq.~\eqref{eq:target_energy_kernel_integrated}. For a given ensemble, we perform the reconstruction for several values of $\lambda$ and $\Delta$. The smearing width $\Delta$ controls how sharp the cut of the timelike spectral density is.
The $\lambda$ parameter, see Eq.~\eqref{eq:def_W_HLT}, is the regulator for statistical noise. For small $\lambda$ the mismatch between the approximated and the target kernels is smaller, but the statistical noise is enhanced, while large $\lambda$ parameters reduce the statistical noise, while enhancing the systematic uncertainty.
 We proceed by scanning over a large range of $\lambda \in [10^{-14},10^{-3}]$ using the identity as a regulator, see Eq.~\eqref{eq:def_W_HLT}.
 In our approach, we tune both parameters to find a balance between systematic and statistical uncertainty.

In addition to $\lambda$ and $\Delta$, one can adjust the energy range where the optimization problem is defined, and in particular the lower end $\omega_0$ in Eq.~\eqref{eq:mismatch_functional}.
If the underlying (finite-volume) spectral density has strong support in the region below the chosen $\omega_0$, larger systematic errors are expected since the approximated kernel differs significantly from the target one for $\omega < \omega_0$, since that region is excluded from the least-squares procedure. 
With the intent of keeping a possibly large $\omega_0$, such a systematic error may however be reduced if the part of the spectrum below $\omega_0$ is known and it is subtracted from the correlator and reconstructed exactly. A similar observation, namely the fact that subtracted correlators de-facto solve the inverse problem up to the subtraction scale was already present in Ref.~\cite{Bruno:2020kyl}. Here, we test in a numerical example and demonstrate that it leads to significant benefits in the reconstructed densities, even for smearing kernels focused more at higher energies.

Hence to reduce the complexity of the reconstruction, we subtract the lowest $N$ states from the correlator defining
\ba 
\label{eq:GWlarger_subtracted}
G^{I=1,\text{sub}}(t) = G^{I=1}(t) - \sum_{n=0}^{N_{\text{rec}}} V_n(L) e^{-E_n(L) t} \,.
\ea
For this subtraction, we use the energies $E_n$ and overlap factors $V_n$ obtained from a previous analysis of the long-distance window quantity \cite{RBC:2024fic}, where several two-pion operators have been employed alongside the GEVP method.
With the GEVP analysis of Ref.~\cite{RBC:2024fic} the number of operators was tuned to reach more or less always the same energy around $1$ GeV. Evidently due to the volume dependence the number of subtracted states varies for each ensemble and a combination of results from different ensembles is possible only at the level of the smeared density, where we have an individual estimate of the systematic uncertainty on every ensemble, according to Eq.~\eqref{eq:syst_uncertainty_calc}.

\begin{figure}
\begin{subfigure}{0.49\textwidth}
\centering
    \includegraphics[width=0.99\linewidth]{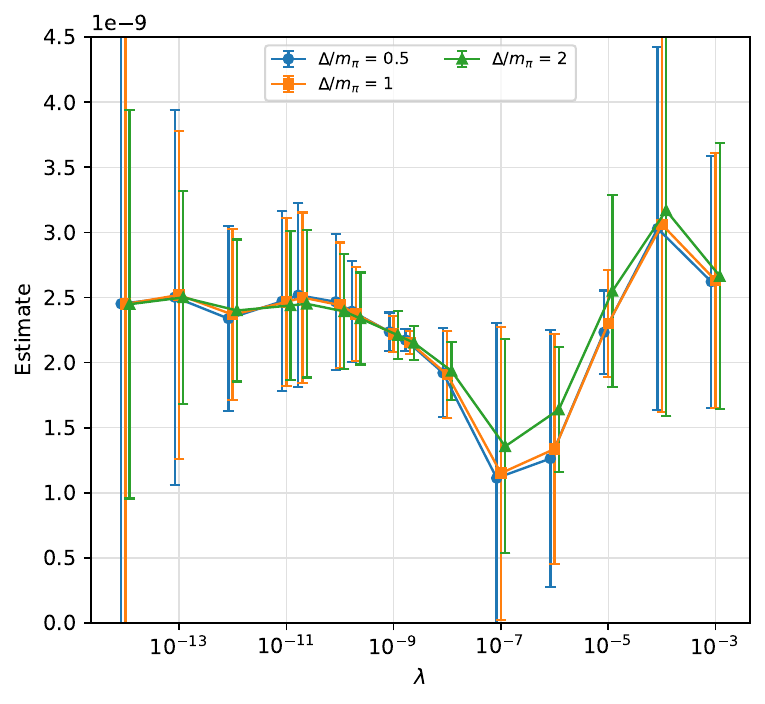}
    \caption{Reconstructed result using $G^{I=1}(t)$ }
    \label{fig:stability_id_nrec=0}
\end{subfigure}
    \begin{subfigure}{0.49\textwidth}
\centering
    \includegraphics[width=0.99\linewidth]{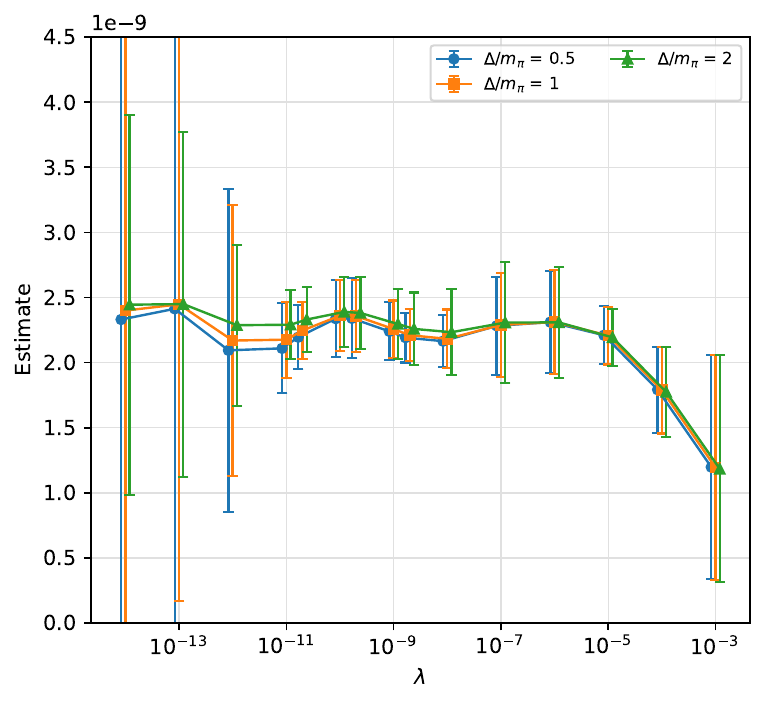}
    \caption{Reconstructed result using $G^{I=1,\text{sub}}(t)$}
    \label{fig:stability_id_nrec=5}
\end{subfigure}
\caption{Stability plot of the integrated result for the intermediate-distance window quantity on the ensemble 48I. The error bars are the combination of statistical and systematic uncertainties.}
\label{fig:stability_id}
\end{figure}

A comparison of the reconstructed result for the intermediate-distance window quantity $a_\mu^{>,\text{win}}(\omega_{\text{cut}}=m_{\tau})$ in the parameter space $[\lambda,\Delta]$ is depicted in Fig.~\ref{fig:stability_id}. For the calculation of $q_{n,\lambda}$ in Eq.\eqref{eq:HLT_recon_awin} we choose the lower bound for the integration in energy-space  $\omega_0=E_{N_{\text{rec}}}$.

Here, we observe that using the subtracted correlator enhances the stability of the reconstruction. This is due to the fact that the approximated kernel $\widehat{T}_{\Delta,\lambda}(\omega)$ does not need to suppress the region below $1$ GeV, if this region is already subtracted from the lattice correlator.

\begin{figure*}
\begin{subfigure}{0.49\textwidth}
\centering
    \includegraphics[width=0.99\linewidth]{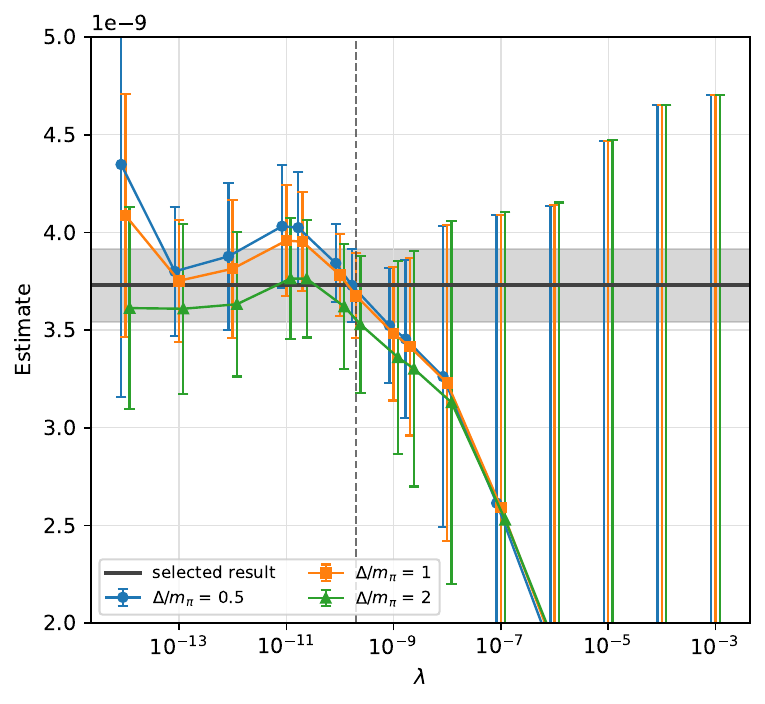}
    \caption{Reconstructed result for $a_\mu^{>,ID}(\omega_{\text{cut}}=1200 \text{ MeV})$ on ensemble 96I}
    \label{fig:stability_id_nrec=0_96I}
\end{subfigure}
    \begin{subfigure}{0.49\textwidth}
\centering
    \includegraphics[width=0.99\linewidth]{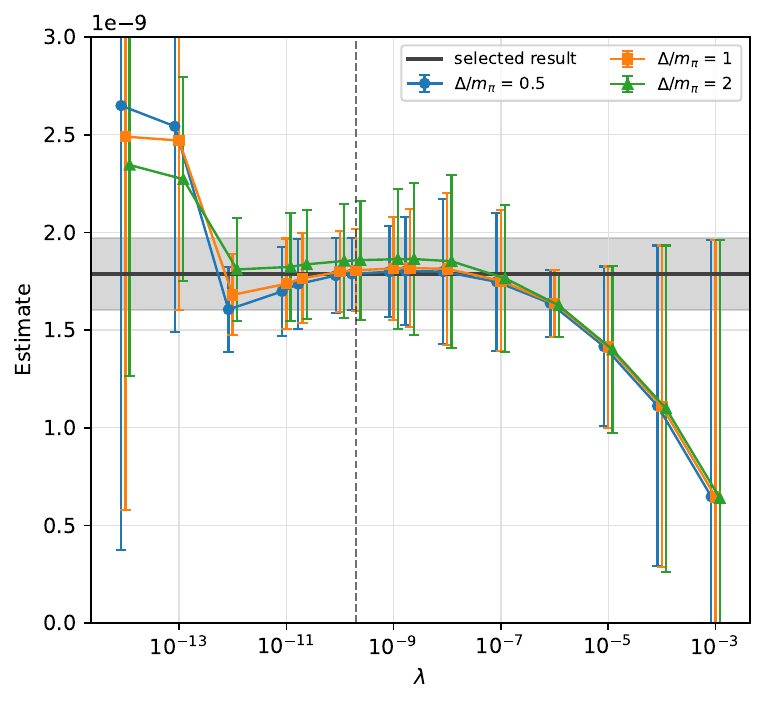}
    \caption{Reconstructed result for $a_\mu^{>,ID}(\omega_{\text{cut}}=m_\tau)$ on ensemble 64I}
    \label{fig:stability_id_nrec=5_96I}
\end{subfigure}

\begin{subfigure}{0.49\textwidth}
\centering
    \includegraphics[width=0.99\linewidth]{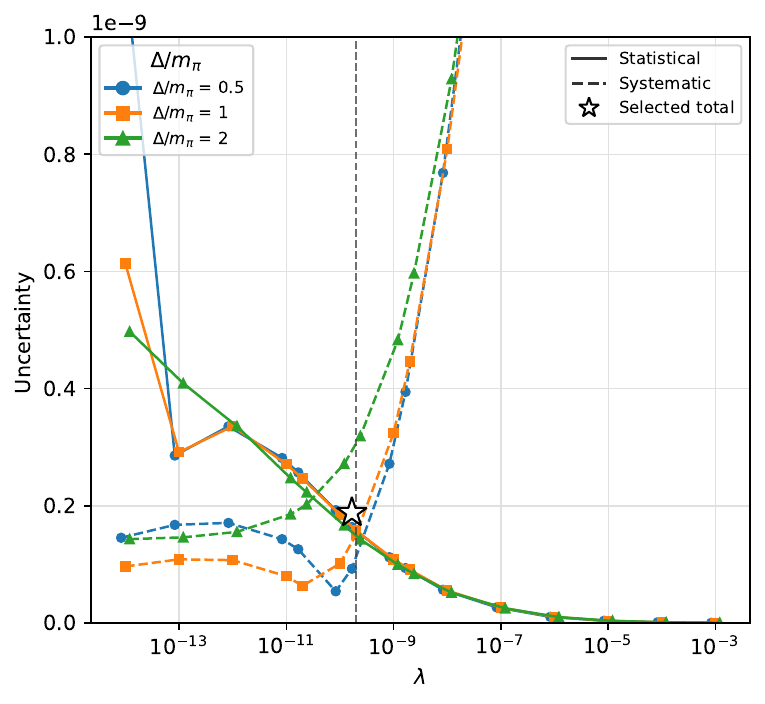}
    \caption{Uncertainties for the reconstruction of $a_\mu^{>,ID}(\omega_{\text{cut}}=1200 \text{ MeV})$ on ensemble 96I}
    \label{fig:stability_id_nrec=0_64I}
\end{subfigure}
    \begin{subfigure}{0.49\textwidth}
\centering
    \includegraphics[width=0.99\linewidth]{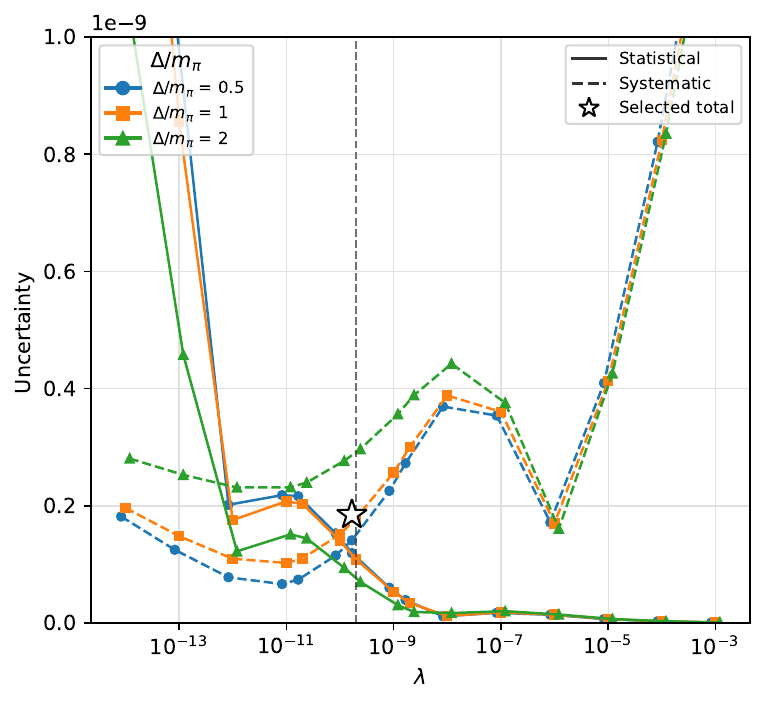}
    \caption{Uncertainties for the reconstruction of $a_\mu^{>,ID}(\omega_{\text{cut}}=m_\tau)$ on ensemble 64I}
    \label{fig:stability_id_nrec=5_64I}
\end{subfigure}
\caption{Top: Plot of the reconstructed results $a_\mu^{>,ID}(\omega_{\text{cut}})$ for scan over $[\lambda,\Delta]$ with total uncertainty. The black horizontal line represents the selected result, with gray error band. Bottom: Comparison of systematic and statistical uncertainties. The star marks the total uncertainty of the selected result.}
\label{fig:selection}
\end{figure*}

In order to select a pair of $\lambda$ and $\Delta$, we investigate the statistical and systematic uncertainty computed via Eqs.~\eqref{eq:GW_stat_unc} and \eqref{eq:syst_uncertainty_calc} respectively. This comparison together with the reconstructed result is depicted in Fig.~\ref{fig:selection}. We observe that the statistical uncertainty increases for small $\lambda$, while the systematic uncertainty tends to increase for large $\lambda$, matching expectations. 
The computed systematic uncertainty according to Eq.~\eqref{eq:syst_uncertainty_calc}, however, does not decrease strictly monotonously, due to oscillations of the approximated kernel around the target kernel, see Fig.~\ref{fig:kernel_mismatch}, which leads to an oscillatory behavior of the systematic uncertainty. Without constraining the  approximated kernel at the endpoint $\omega_0$ to be zero exactly, see Eq. \eqref{eq:q_coef_endpoint_constraint}, this behavior is even more severe, which may cause additional difficulties in picking a conservative error estimate. 

To obtain a single result for each ensemble, we select the minimum of the total uncertainty in the parameter space $[\lambda,\Delta]$ and for all results selected in this way, we find that statistical and systematic errors are typically well balanced.
The observed oscillation of the systematic uncertainty is due to the fact that the approximated kernel $\widehat{T}_{\Delta,\lambda}(\omega)$ oscillates around the sharp target kernel $T(\omega)$. 

Although we focus on the intermediate-distance window in this work, a similar stability is expected for the short-distance and long-distance window quantities. For the long-distance window quantity, we want to mention here that it may be more advantageous for the stability of the reconstruction to use every second timeslice in the basis functions.

\section{Analysis of the reconstructed observable}
\label{sec:analysis_lat}
\begin{figure}
    \centering
    \includegraphics[width=0.99\linewidth]{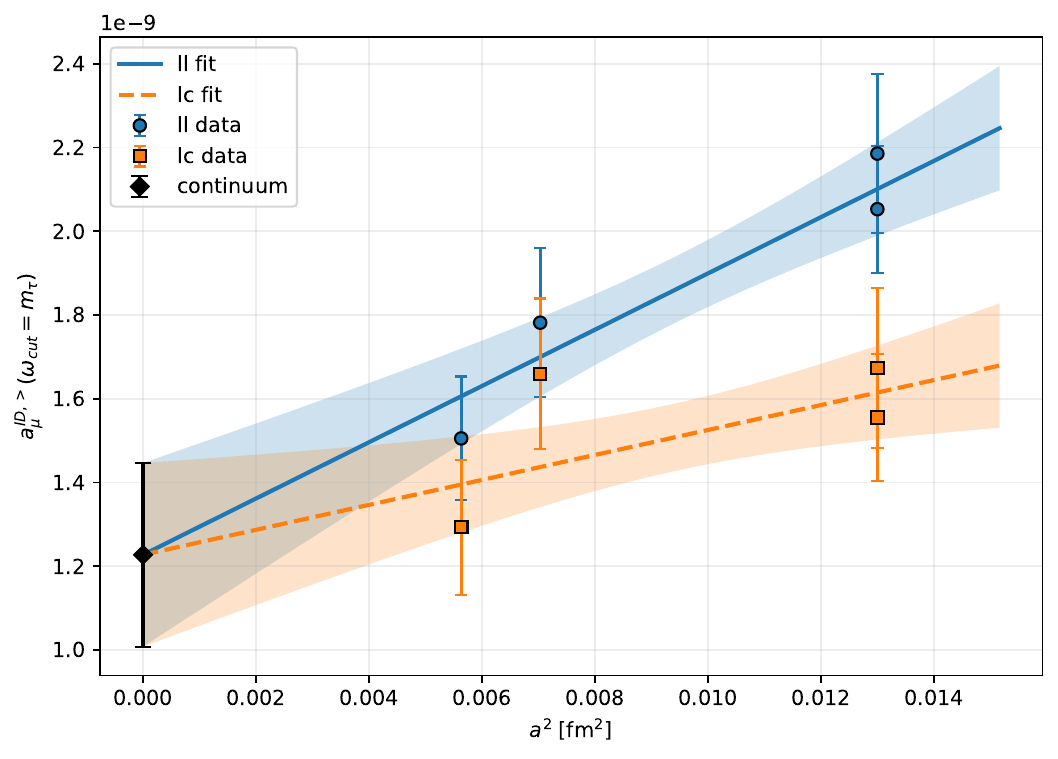}
    \caption{Constrained fit  -- with the largest weight in the AIC model average -- to the lc and ll data without finite volume correction for the continuum extrapolation of the result for $a_\mu^{>,ID}(\omega_{\text{cut}}=m_\tau)$.}
    \label{fig:continuum_ID}
\end{figure}
Following the procedure from section \ref{sec:recon_lat}, we obtain a result for $a_\mu^{>,\text{ID}}(\omega_{\text{cut}})$ for each ensemble with a statistical and systematic uncertainty from the reconstruction. We perform the reconstruction for lattice QCD correlators computed with two different discretizations of the vector current, one version where both vector insertions in Eq.\eqref{eq:spatsummedcor} are local vector currents (ll) and a variant, where one vector current is local and the other one is a conserved vector current (lc).
To extrapolate to the continuum, we consider the fit function
\ba 
f(a^2,m_\pi L) = b_0+b_1 a^2 + b_2 e^{-m_\pi L}
\ea 
and perform a model averaging using the AIC criterion \cite{Akaike:1974vps}. In the average, we include the fits to the ll or the lc separately, but also constrained fits where the parameter $b_0$ is shared. We also consider models, without finite volume correction, thus neglecting the term with $b_2$. In this study, we do not account for the small pion mass mistuning between the ensembles. For the intermediate-distance window this effect is expected to be below the achieved precision of the reconstructed result. However, for the long-distance window quantity this effect may not be negligible anymore.
A combined fit to the ll and lc data is shown in Fig.~\ref{fig:continuum_ID}.
For the final result, we assume $\sigma_{\text{total}}^2=\sigma^2_{\text{stat}}+\sigma^2_{\text{syst}}$, where the systematic uncertainty is due to the reconstruction. We note that, since we are using a very conservative upper bound for the systematic uncertainty improving on this bound and thus reducing the systematic uncertainty, a smaller statistical uncertainty may be achieved in the balancing procedure. In order to separate the systematic and statistical uncertainties for the model averaged result, we perform the same model averaging procedure in a scenario, where the systematic uncertainty is artificially increased by a factor two. By comparing both scenarios and assuming that the statistical uncertainty is unchanged, we compute the contribution of the systematic uncertainty to the final result.

\begin{table}[]
\caption{Comparison of reconstructed continuum-extrapolated lattice QCD result with perturbative QCD. The pQCD results are obtained in the same way as in Table \ref{tab:pheno_and_pertQCD}. The pQCD results for $\omega_{\text{cut}}=1400, 1600$ MeV are only given for comparison and may underestimate the systematic uncertainty from truncating the perturbative series. For $\omega_{\text{cut}}=1200$ MeV, the procedure does not yield a robust pQCD estimate. Results are given in units of $10^{-10}$.
}
\label{tab:continuum_reconstructed}
\begin{tabular}{|c|c|c|}\hline
 & \multicolumn{2}{|c|}{$a_\mu^{>,\text{ID}}(\omega_{\text{cut}})$  }\\\hline

$\omega_{\text{cut}} $ [MeV] & lattice QCD &pQCD (4th order)\\\hline
1200 & $36.1(3.4)_{\mathrm{stat}}(1.5)_{\mathrm{syst}}$ & \\ \hline
1400 & $27.8(1.5)_{\mathrm{stat}}(1.3)_{\mathrm{syst}}$ &  $26.04     [^{+0.34}_{-3.59}]$  \\ \hline
1600 & $18.0(2.3)_{\mathrm{stat}}(1.3)_{\mathrm{syst}}$ &  $17.97     [^{+0.19}_{-0.77}]$\\ \hline
1776.86 & $11.5(2.4)_{\mathrm{stat}}(1.4)_{\mathrm{syst}}$ & $13.05     [^{+0.12}_{-0.31}]$  \\ \hline
2000 & $8.64(1.2)_{\mathrm{stat}}(0.66)_{\mathrm{syst}}$ & $8.82     [^{+0.07}_{-0.12}]$  \\ 
\hline

\end{tabular}
\end{table}

\section{Discussion of the results}
\label{sec:discussion}
In this work, we have investigated the hadronic vacuum polarization contribution above a timelike cut $\omega_{\text{cut}}$. For the fully inclusive determination of the HVP from $\tau$ decays, the case of $\omega_{\text{cut}}=m_\tau$ is special, since it represents the upper end of the spectrum that is available from these decays. For having better control over systematic effects due to the experimental measurements, it might be benericial, however, to substitute the part above a smaller value of $\omega_{\text{cut}}$. By balancing experimental and theoretical uncertainties, it might be even advantageous to consider a smeared -- instead of a sharp -- cut in energy.

The observable $a_\mu^{>}(\omega_{\text{cut}})$ from lattice QCD furthermore presents an interesting case of an inverse problem, where the form of the underlying spectral density is known up to some degree. By using this knowledge, it is possible to calculate a concrete systematic error of the reconstruction, while the central value of the reconstructed observable is a lattice QCD result.
Our main result for the intermediate-distance HVP contribution above $\omega_{cut}=m_\tau$ is 
\ba
a_\mu^{>,\text{ID}}(\omega_{\text{cut}}=m_\tau) &= 11.5(2.4)_{\mathrm{stat}}(1.4)_{\mathrm{syst}}  \cdot 10^{-10}
\ea
This result, as well as the results for $\omega_{\text{cut}}=\{1400, 1600,2000\}$ MeV are in good agreement with the perturbative QCD computation, see Table \ref{tab:continuum_reconstructed}. Especially for the lower values of $\omega_{\text{cut}}$ this finding is somewhat surprising, since this region is below the threshold, where perturbative QCD can be safely applied.
With the current precision of the lattice result, isospin breaking effects which are typically of order $1\%$ are negligible.

The uncertainty of the lattice result is much higher than the pQCD theory band error for $\omega_{\text{cut}}=m_\tau$. However, for a fully inclusive determination of the HVP from $\tau$-data, as described in Ref.~\cite{Bruno:2026yba}, our lattice calculation now offers the possibility to supplement the contribution $a_\mu^{>,\text{ID}}$ for a lower cut in the energy spectrum. We note that for the pQCD results for $\omega_{\text{cut}}=1400, 1600$ in Table \ref{tab:continuum_reconstructed}, we use again 4th order pQCD \footnote{As in Table \ref{tab:pheno_and_pertQCD}, the uncertainties are calculated from varying the renormalization scale $\mu \in [\sqrt{s}/2,2\sqrt{s}]$ and the strong coupling at the Z-pole mass $\alpha_s = 0.118\pm 0.003$}. Although this may still underestimate the true systematic uncertainty from truncating the perturbative series, the agreement between lattice and pQCD may hint to the fact that duality violations largely cancel in this particular observable.

In this work, we do not investigate $\omega_{\text{cut}}<1$ GeV. But, we note that these can be investigated by considering instead of the subtracted correlator \eqref{eq:GWlarger_subtracted}, the complementary part of the correlator.
Even though we only present results for the intermediate-distance window quantity, a preliminary study of the long-distance window quantity on all ensembles considered in this work shows a reconstructed result of $a_\mu^{>,\text{LD}}(\omega_{\text{cut}}=m_\tau)<1 \cdot 10^{-10}$.

We expect the method to be applicable for the short-distance window as well. This quantity however, may be even more affected by cutoff effects than the short-distance window evaluated without a timelike cut. In order to perform a continuum extrapolation it may therefore be necessary to include ensembles with even finer lattice spacing.

Finally, we would like to mention here again that the authors of Ref.~\cite{Allen:2026iad} found a good agreement  of the short-distance window observable between lattice QCD and perturbative QCD, when applied for energies well below $m_\tau$. Through further studies of this quantity using pQCD and the reconstruction methods given in the present manuscript, a deeper understanding of the quark-hadron duality may be reached.

\section{Acknowledgments}
The authors would like to thank Maarten Golterman and Santiago Peris for helpful discussions.
We furthermore thank our colleagues of the RBC and UKQCD collaborations for many valuable discussions and joint efforts over the years.
H.-Y. W. is supported by the China Scholarship Council (No. 202608080246).
The authors gratefully acknowledge the Gauss Centre for Supercomputing
e.V. (www.gauss-centre.eu) for funding this project by providing
computing time on the GCS Supercomputer JUWELS at Jülich
Supercomputing Centre (JSC).  We acknowledge the EuroHPC Joint
Undertaking for awarding this project access to the EuroHPC
supercomputer LUMI, hosted by CSC (Finland) and the LUMI consortium
through a EuroHPC Extreme Scale Access call as well as the EuroHPC
supercomputer LEONARDO, hosted by CINECA (Italy).  An award of
computer time was provided by the ASCR Leadership Computing Challenge
(ALCC) and Innovative and Novel Computational Impact on Theory and
Experiment (INCITE) programs. This research used resources of the
Argonne Leadership Computing Facility, which is a DOE Office of
Science User Facility supported under contract DE-AC02-06CH11357. This
research also used resources of the Oak Ridge Leadership Computing
Facility, which is a DOE Office of Science User Facility supported
under Contract DE-AC05-00OR22725.  This research used resources of the
National Energy Research Scientific Computing Center (NERSC), a
U.S. Department of Energy Office of Science User Facility located at
Lawrence Berkeley National Laboratory, operated under Contract
No.~DE-AC02-05CH11231 using NERSC award NESAP m1759 for 2020. This
work used the DiRAC Blue Gene Q Shared Petaflop system at the
University of Edinburgh, operated by the Edinburgh Parallel Computing
Centre on behalf of the STFC DiRAC HPC Facility
(www.dirac.ac.uk). This equipment was funded by BIS National
E-infrastructure capital grant ST/K000411/1, STFC capital grant
ST/H008845/1, and STFC DiRAC Operations grants ST/K005804/1 and
ST/K005790/1. DiRAC is part of the National E-Infrastructure.  We
gratefully acknowledge disk and tape storage provided by USQCD and by
the University of Regensburg with support from the DFG.
The lattice data analyzed in this project was generated using GPT
\cite{GPT} and Grid \cite{GRID,Boyle:2016lbp,Yamaguchi:2022feu}. The reconstruction of the target quantity was performed using SpectRAL \cite{SpectRAL} .
 The authors acknowledge the use of GPT Astra and Claude Opus 5.5 for code generation and spell checking. All final outputs were manually
verified and remain the sole responsibility of the authors.

\appendix

\section{Details of the implementation of the HLT algorithm}
\label{sec:appendix_HLTimplementation}

The coefficients \(q_n\) are determined by minimizing the squared distance between the reconstructed kernel and the target kernel. More generally, the lower limit of the mismatch integral may be chosen as a physical spectral threshold $\omega_0\geq 0$;
the mismatch functional is
\ba
\label{eq:mismatch_functional}
A[q]
=
\int_{\omega_0}^\infty d\omega
\left[
\widehat{T}_{\Delta,\lambda}(\omega)
-
T_\Delta(\omega)
\right]^2.
\ea
By minimizing $A[q]$ one obtains the algebraic equation 
\ba
\sum_n A_{nm} \, q_n = b_m
\ea
involving the matrix 
\ba
A_{nm}
=
\int_{\omega_0}^\infty d\omega \,
\phi_{t_n}(\omega) \,
\phi_{t_m}(\omega)\,,
\ea
and the vector
\ba
b_n
=
\int_{\omega_0}^\infty d\omega \,
\phi_{t_n}(\omega) \,
T_\Delta(\omega).
\ea
The inversion of $A_{nm}$ is regulated by the identity operator~\cite{Bruno:2024} rather than the covariance matrix and, following the HLT prescription~\cite{Hansen:2019idp}, the coefficients are obtained by minimizing the combination
\ba
\label{eq:def_W_HLT}
\mathcal W_\lambda[q]
=
(1-\lambda)A[q]
+
\lambda B[q],
\qquad 0\leq\lambda\leq 1\,.
\ea
 where we use the identity as the regulator for the statistical noise, $B[q]=q^T q$, ~\cite{Bruno:2024fqc},
 in contrast to the standard choice 
$B[q]=\frac{q^T\mathrm{Cov}\,q}{G(0)^2}$. 
To compensate for the presence of $(1-\lambda)$ the r.h.s.~of the normal equation is adapted to
\ba
f_n(\lambda)=(1-\lambda)b_n.
\ea
and the coefficients acquire a natural dependence on $\lambda$, thereby justifying the notation $q_{n,\lambda}$.
Similarly to the area constraint of the original HLT algorithm \cite{Hansen:2019idp}, one can constrain the kernel to a fixed value $d$ at a given energy $\omega_f$
\ba 
v_n = \phi_{t_n}(\omega_f), \qquad v^T q = d 
\ea 
The constrained system is
\ba
W(\lambda) q+ \mu v = f(\lambda),
\ea
where $\mu$ is a Lagrange multiplier.
The solution for the coefficients $q$ is given by 
\ba
\label{eq:q_coef_endpoint_constraint}
q
=
W(\lambda)^{-1} f(\lambda)
+
W(\lambda)^{-1}v
\frac{
d-v^T W(\lambda)^{-1}f(\lambda)
}{
v^T W(\lambda)^{-1}v
}.
\ea 
For the approximation of the kernel in Eq.~\eqref{eq:target_energy_kernel_integrated} we constrain the approximated kernel to be zero at the endpoint, e.g. $d=0$ for $\omega_f=\omega_0$.

\bibliography{references}

\end{document}